\documentclass[9pt,twocolumn,twoside]{opticajnl}

\usepackage{stfloats}
\usepackage{colorspace}
\definecolorspace*{sRGB}{iccbased}{sRGB.icc}
\pagecolorspace{sRGB}

\journal{opticajournal} 

\setboolean{shortarticle}{false}

\usepackage{lineno}

\title{Fresnel lens time-to-space encoding for single-shot ultrabroadband terahertz spectroscopy}

\author[1,2]{Alexander Holm Ohrt}
\author[2]{Man Tou Wong}
\author[1]{William Vang Carstensen}
\author[1]{Long Cheng}
\author[2]{In Hyeok Choi}
\author[2]{Keith Adam Nelson}
\author[1]{Binbin Zhou}
\author[1,*]{Peter Uhd Jepsen}

\affil[1]{Department of Electrical and Photonics Engineering, Technical University of Denmark, 2800 Kongens Lyngby, Denmark}
\affil[2]{Department of Chemistry, Massachusetts Institute of Technology, Cambridge, Massachusetts 02139, USA}

\affil[*]{puje@dtu.dk}

\begin{abstract}
Single-shot terahertz waveform detection captures irreversible and shot-dependent phenomena inaccessible to repeated delay scanning. Most established schemes use approximately linear temporal encoding, so a fixed number of resolved spatial points imposes a trade-off between temporal sampling and observation window. Here, we demonstrate the use of a Fresnel lens to produce controlled, approximately quadratic time-to-space encoding for single-shot terahertz detection. The calibrated nonuniform mapping spans 2.4 ps with local groove-defined temporal increments down to 12 fs. Combined with coherent air-based nonlinear detection, the encoder enables field-resolved single-shot measurements of freely propagating THz pulses from a two-color air plasma driven by a tabletop 1-kHz Ti:sapphire amplifier. The reconstructed waveforms contain spectral content beyond 10 THz and agree with conventional delay scans. We further demonstrate transmission-mode single-shot terahertz time-domain spectroscopy of silicon and shot-to-shot monitoring of droplet-induced perturbations to the air-plasma source. These results show that a simple, low-cost Fresnel lens can provide nonuniform temporal encoding that combines high temporal resolution with a comparatively large observation window.
\end{abstract}

\setboolean{displaycopyright}{false} 

\doi{}
\begin{document}

\maketitle

\section{Introduction}

Conventional terahertz time-domain spectroscopy (THz-TDS) reconstructs the electric-field waveform of a THz pulse by repetitive sampling at successive probe–THz delays, requiring a stable source and full recovery of the sample to equilibrium between successive excitations. Single-shot waveform detection instead records a range of delays in parallel, allowing complete waveforms to be acquired from individual events \cite{Teo2015SingleShotReview}. More broadly, single-shot ultrafast measurements have revealed rare phenomena obscured by averaging, including optical rogue waves \cite{Solli2007OpticalRogueWaves}, and enabled studies of nonrecovering metastable phases \cite{Gao2022MetastableHiddenPhase} and material responses above permanent-damage thresholds \cite{Wakeham2000DualEchelon}. In THz experiments, eliminating the waveform scan also enables measurements of fluctuating, irreversible, and shot-dependent processes and makes multidimensional spectroscopy more practical \cite{Gao2022EchelonBalanced,Zhang2024MagnonUpconversion,Zhang2024MagnonNonlinearCoupling}. Field-resolved bandwidth is equally important because the highest-frequency components are particularly sensitive to the fastest current and polarization dynamics \cite{Leitenstorfer1999ChargeTransport,Cooke2012MobileChargeGeneration,Dringoli2024SnSePhaseChange}. A need therefore remains for a tabletop method that combines single-shot field acquisition with ultrabroadband sensitivity.

Here, we introduce a time-to-space encoding scheme based on a refractive cylindrical Fresnel lens, using its controlled, approximately quadratic group-delay profile for single-shot THz waveform detection. The lens focuses the optical probe, while its grooved profile removes material that would otherwise compensate transverse arrival-time differences in a conventional lens. Probe light from different transverse positions therefore reaches the THz focus at different times. For a fixed aperture and number of resolved spatial points, the nonuniform mapping spans a longer time window than a linear mapping with the same smallest local temporal increment. The balance between local temporal increments and the observation window can be adjusted simply by translating the lens laterally relative to the probe, thereby selecting a different part of its aperture. The resulting nonuniform temporal axis is determined experimentally and fully accounted for during waveform reconstruction.

Existing single-shot THz methods encode delay in the probe spectrum, across the probe beam, or through a noncollinear interaction geometry. Spectral encoding with chirped probes supports rapid acquisition, but the finite probe bandwidth must represent the complete temporal record, limiting the simultaneous observation window and temporal resolution \cite{Jiang1998ChirpedOpticalBeam,Roussel2022PhaseDiversityEOS,Nallapareddy2025SETS,Couture2023SinglePulse50kHz}. Reflective and transmissive echelons impose an approximately linear spatial sequence of probe delays and have enabled high-speed multidimensional THz spectroscopy \cite{Gao2022EchelonBalanced,Zhang2024MagnonUpconversion,Zhang2024MagnonNonlinearCoupling} and scanning-free optical and THz Kerr-effect measurements \cite{Mead2019EchelonKerr}. More recently, the same encoding principle was implemented using a blazed reflection grating, which supported waveform detection beyond 6 THz with a 35-fs probe \cite{Wu2026ReflectionGratings}. For a finite aperture, however, maintaining a fine constant temporal increment throughout a linear mapping restricts the total observation window. Curved-wavefront encoding creates a nonlinear radial time-to-space mapping by deliberately defocusing the probe at the electro-optic crystal, so that different probe radii sample the THz pulse at different delays. Its demonstrated bandwidth remained below 2.5 THz, and because the delay coordinate is distributed directly across the interaction area, the usable time window is limited by the transverse overlap of the defocused probe and THz beams \cite{Saxena2018CurvedWavefront}. Noncollinear encoding has achieved scanning-free detection beyond 30 THz, although tabletop operation required averaging, while single-shot operation at comparable bandwidth required millijoule-level THz pulses \cite{Shan2000NoncollinearEOS,Ohrt2025ScanningFree,Ohrt2025SingleShotHighEnergy}.

We combine the Fresnel lens encoder with coherent balanced air-based nonlinear THz detection, avoiding the absorption and phase-matching constraints of electro-optic crystals \cite{Dai2006AirPlasmaDetection,Ohrt2024BalancedAirBiased,Huang2025AirAidedCoherentDetection}. Earlier coherent air-based single-shot measurements probed THz-correlated second-harmonic generation directly within the emitting air-plasma filament, a source-local geometry without a separate THz path for sample interaction \cite{GarrigaFrancis2023LocalTFISH,Chong2024SingleShotLocalTFISH}. Here, the encoded probe and freely propagating THz field interact at a downstream focus, leaving an intermediate sample plane for THz-TDS. The calibrated mapping spans 2.4 ps with local groove-defined temporal increments down to 12 fs. Using 100-fs pulses from a tabletop 1-kHz Ti:sapphire amplifier, we reconstruct two-color air-plasma THz waveforms with spectral content beyond 10 THz. The waveforms agree with conventional delay scans. We further demonstrate transmission-mode single-shot THz-TDS of silicon and shot-to-shot monitoring of droplet-induced perturbations to the THz source. Together, these results extend single-shot THz-TDS to coherent air-based detection of two-color air-plasma THz pulses.

\section{Results and discussion}

\subsection{Principle of Fresnel lens time-to-space encoding}

\begin{figure}[H]
    \centering
    \includegraphics[width=0.45\textwidth]{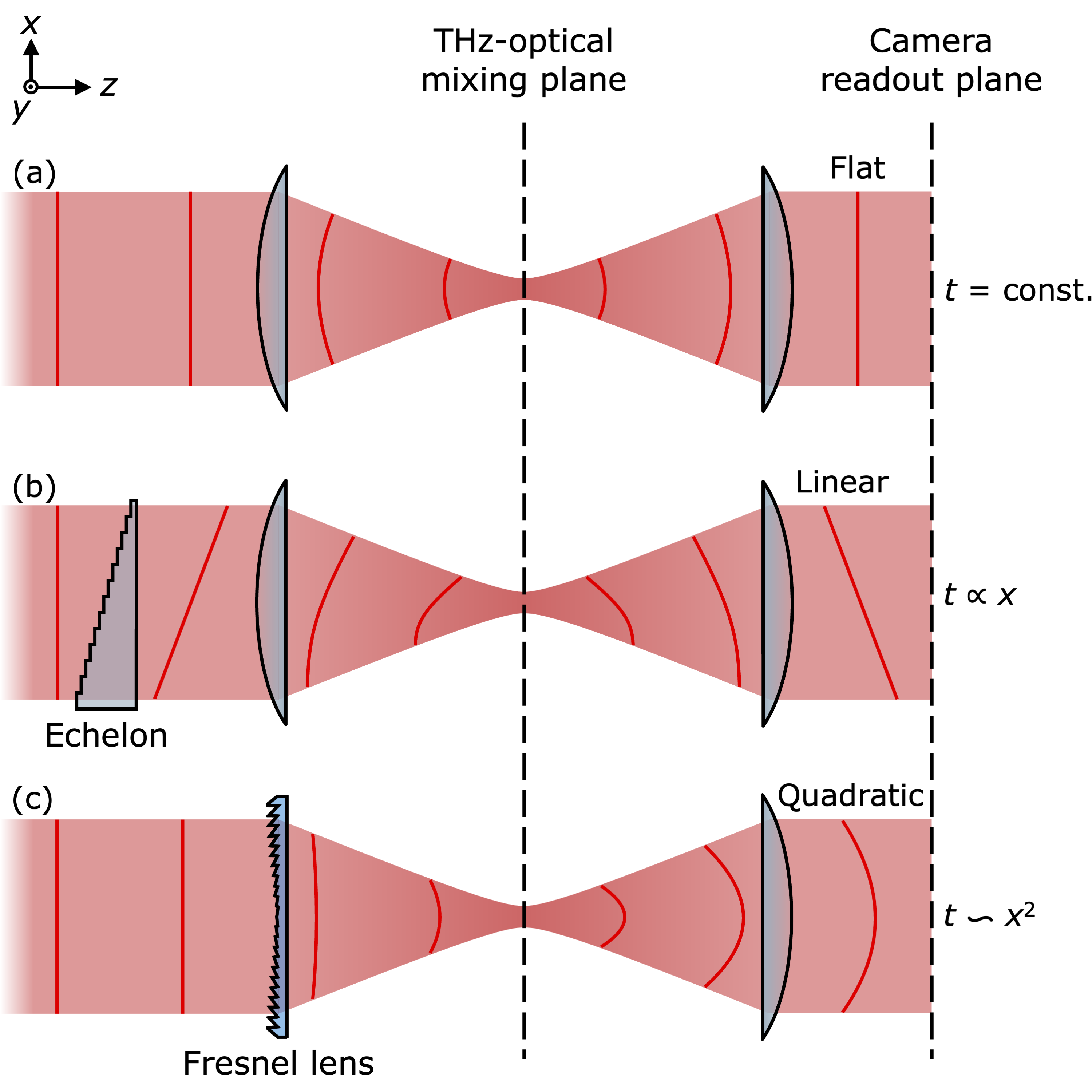}
    \caption{Conceptual evolution of the probe pulse front through different imaging systems. Red curves show successive snapshots of the probe pulse front during propagation, while the shaded red regions indicate the spatial extent of the probe beam. (a) An ideal conventional imaging system restores a flat pulse front at the camera plane, giving a spatially uniform arrival time, $t=\mathrm{const.}$. (b) A transmissive echelon introduces successive transverse delays whose envelope produces an approximately linear time-to-space mapping, $t(x)\propto x$. (c) A cylindrical Fresnel lens produces a curved arrival-time front corresponding to an approximately quadratic mapping, $t(x)\sim x^2$.}
    \label{fig:1}
\end{figure}

The operating principle can be understood by considering the evolution of an optical probe pulse front in Fig.~\ref{fig:1}. Figure~\ref{fig:1}(a) shows an ideal conventional imaging system in which a collimated probe is focused and subsequently recollimated while its spatially uniform arrival time is restored at the camera plane. Although marginal rays travel farther through air than axial rays, they traverse less lens material. The resulting material and geometric delays compensate such that the optical path from the incident pulse front to the focus is equal for all rays, consistent with Fermat's principle of stationary optical path. Consequently, all transverse positions in the recollimated beam correspond to the same probe–THz delay, $t=\mathrm{const.}$, and a spatially resolved detector records only one point of the THz waveform.

Time-to-space encoders deliberately break this equal-arrival-time condition. A transmissive or reflective echelon introduces successive optical-path delays across the probe, producing spatially separated beamlets whose arrival times follow an approximately linear envelope, as illustrated in Fig.~\ref{fig:1}(b). At the common probe–THz focus, each beamlet samples a different relative delay. Imaging the probe onto a spatially resolved detector then converts the transverse coordinate into a temporal coordinate. For a constant echelon step depth, this produces a discrete, approximately linear mapping, $t(x)\propto x$, where $x$ denotes the transverse coordinate along the encoding direction.

A Fresnel lens provides a simple way to retain focusing while introducing a nonlinear transverse group-delay profile, as illustrated in Fig.~\ref{fig:1}(c). It replaces the continuous thickness profile of a conventional lens with repeated facets, removing sections of material that would otherwise provide continuous group-delay compensation. The facets still direct the probe toward a common focus, but pulse envelopes originating from different transverse positions arrive there at different times. The optical phase determines the focusing, whereas the group delay determines the pulse arrival time and therefore the time-to-space encoding. A cylindrical Fresnel lens confines the encoding to one transverse coordinate, $x$, leaving the orthogonal coordinate, $y$, available for independent focusing.

A simple geometric model captures the dominant Fresnel-induced mapping. Consider a point on the lens at transverse coordinate $x$, measured from the lens axis in the focusing direction. The free-space distance from the nominal lens plane at $x$ to the focus is $\sqrt{f^2+x^2}$, whereas the on-axis distance is the Fresnel lens focal length $f$. The same geometrical path difference is present in the other systems in Fig.~\ref{fig:1}, but it is not their complete group-delay mapping: the continuous material-thickness profile of a conventional lens compensates it, while an echelon deliberately adds a step-dependent optical-path delay on top of this compensation. For the Fresnel lens, we approximate the smooth delay envelope by the free-space contribution and neglect residual position-dependent material delay within the grooves. Taking the on-axis arrival time as zero gives

\begin{equation}
    t(x)=\frac{\sqrt{f^2+x^2}-f}{c},
    \label{eq:time_window}
\end{equation}

where $c$ is the speed of light. In the paraxial limit, $|x|\ll f$, the leading-order term is

\begin{equation}
    t(x)\simeq\frac{x^2}{2fc},
\end{equation}

which gives the approximately quadratic time-to-space mapping. Experimental support for the model is presented later in Fig.~\ref{fig:4}(a) and Supplementary Figs.~S1 and S6(a). The non-paraxial expression in Eq.~(\ref{eq:time_window}) is used for the subsequent calibration, while the quadratic form provides the leading-order interpretation.

The continuous profile $t(x)$ is sampled by the Fresnel grooves. Considering one side of the lens, the groove positions can be written as $x_n=x_0+ng$, where $g$ is the constant groove pitch, $x_0$ denotes the beginning of the illuminated region, and $n$ is the groove index. The local groove-defined temporal increment is therefore

\begin{equation}
    \Delta t(x_n)=t(x_n+g)-t(x_n)
    =\frac{\sqrt{f^2+(x_n+g)^2}-\sqrt{f^2+x_n^2}}{c}.
    \label{eq:time_resolution}
\end{equation}

Using the paraxial expression for $t(x)$, this finite difference becomes

\begin{equation}
    \Delta t(x_n)\simeq
    \frac{(x_n+g)^2-x_n^2}{2fc}
    =\frac{g x_n}{fc}+\frac{g^2}{2fc}.
\end{equation}

Except within the first few grooves adjacent to the lens axis, $x_n\gg g$, and the term proportional to $g^2$ is small. Neglecting this term, the local temporal increment is therefore

\begin{equation}
    \Delta t(x_n)\simeq\frac{g x_n}{fc}.
\end{equation}

\begin{figure}[H]
    \centering
    \includegraphics[width=0.45\textwidth]{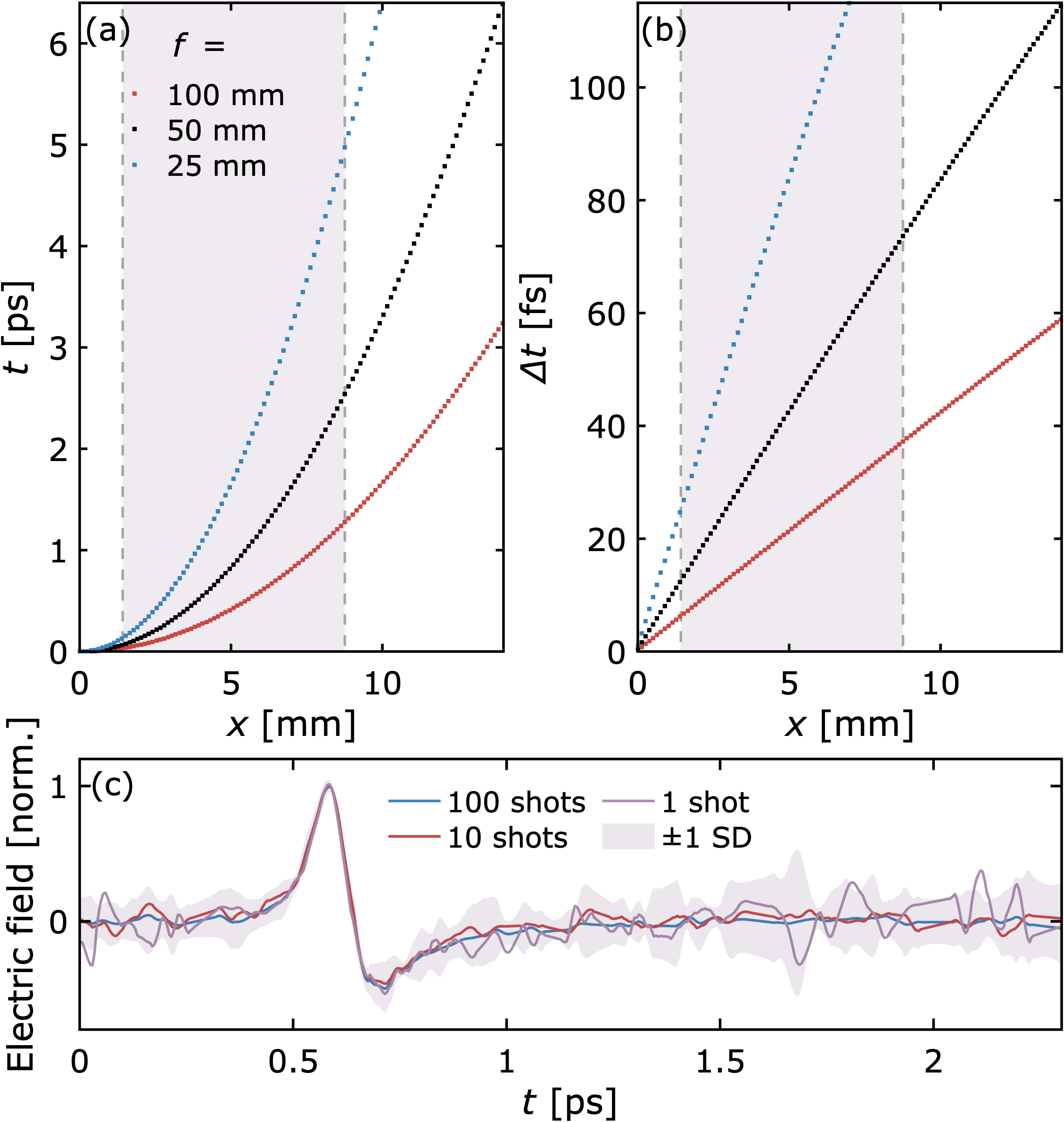}
    \caption{Fresnel lens encoding characteristics and single-shot waveform acquisition. (a) Calculated relative arrival time $t(x)$ for cylindrical Fresnel lenses with focal lengths of 100, 50, and 25 mm, obtained from Eq.~(\ref{eq:time_window}). The shaded region marks the calibrated Fresnel-plane interval used with the approximately 50-mm lens. (b) Corresponding local groove-defined temporal increment, $\Delta t=t(x+g)-t(x)$, calculated from Eq.~(\ref{eq:time_resolution}) using $g=127$ µm. The selected interval provides a 2.42-ps time window with local increments of approximately 12–72 fs. (c) Normalized THz waveforms reconstructed from averages of 100 and 10 acquisitions and from one representative single acquisition. The shaded region denotes $\pm 1$ standard deviation over 200 independently reconstructed single-shot waveforms.}
    \label{fig:2}
\end{figure}

Thus, the arrival time varies approximately quadratically with $x$, while the groove-defined temporal increment increases approximately linearly. The non-paraxial expressions in Eqs.~(\ref{eq:time_window}) and (\ref{eq:time_resolution}) are used for the arrival-time and groove-increment calculations in Fig.~\ref{fig:2}.

Figures~\ref{fig:2}(a) and \ref{fig:2}(b) show the calculated arrival time and local temporal increment for cylindrical Fresnel lenses with different focal lengths. The experimental encoder is a refractive cylindrical acrylic Fresnel lens with an approximately 50-mm nominal focal length and a groove density of 200 grooves/in., corresponding to $g=127$ µm (item no. 80, Fresnel Technologies, Inc.) \cite{FresnelTechnologies2014Catalog}. Its response is represented by the black curves.

Because $t(x)$ is symmetric about the lens axis, $t(-x)=t(x)$, illuminating both sides would encode each delay at two transverse positions. The immediate vicinity of the axis is also less useful experimentally. At $x=0$, the delay gradient vanishes, and the first groove-to-groove increment is approximately $g^2/(2fc)$, corresponding to only about 0.54 fs for the lens used here. The very small delay variation around the axis makes calibration and waveform reconstruction particularly sensitive to spatial averaging and experimental imperfections. Thus, the smallest nominal temporal increment is not necessarily the most useful experimentally.

We therefore displace the Fresnel lens laterally relative to the probe so that the illuminated interval lies entirely on one side of the lens axis. This gives a monotonic time-to-space mapping without duplicate temporal encoding and allows the operating interval to be selected according to the desired balance between local temporal increment and total time window. The shaded region in Figs.~\ref{fig:2}(a) and \ref{fig:2}(b) denotes the Fresnel-plane interval inferred from the experimental calibration. Across the employed interval, approximately 58 groove spacings encode 2.42 ps, corresponding to a mean groove-defined temporal increment of approximately 42 fs. The local increment increases from 12 to 72 fs, with the principal THz transient positioned within the more finely sampled part of the window. The increasing increment should be considered for long-lived, high-frequency coherent oscillations such as magnons or phonons, because coarse sampling later in the window could undersample the oscillation and bias its recovered amplitude or decay time. This can be mitigated by acquiring a small number of single-shot images at different delay-stage positions, thereby shifting the oscillation across the nonuniform mapping and providing interleaved temporal samples. Illumination centered on the lens, revealing both sides of the approximately quadratic profile, is shown in Supplementary Fig.~S1. Supplementary Fig.~S2 estimates the chromatic spreading of the optical probe at the probe–THz mixing plane arising from the Fresnel grooves. Using the treatment adapted from Refs.~\cite{Wu2026ReflectionGratings,Wu2026ReflectionGratingsSupplement}, the estimated spreading is approximately 18–113 µm across the employed interval. Under the present conditions, the reconstructed spectra show no appreciable position-dependent loss of bandwidth across the calibrated interval, as shown in Supplementary Fig.~S6(c). Chromatic spreading should nevertheless be reassessed when shorter, spectrally broader probe pulses are used to extend the detectable THz bandwidth.

Figure~\ref{fig:2}(c) demonstrates waveform reconstruction using the encoded interval. The principal THz transient is recovered from a representative single acquisition, while averaging 10 or 100 acquisitions progressively suppresses shot-to-shot noise. The shaded region shows the standard deviation obtained from 200 independently reconstructed single-shot waveforms. The experimental implementation and calibration of the mapping are described in the following subsection.

\subsection{Experimental implementation and calibration}

In principle, Fresnel lens time-to-space encoding can be combined with different coherent THz detection schemes, including electro-optic sampling (EOS). Here, we choose air-based nonlinear detection because it avoids the absorption and phase-matching constraints of electro-optic crystals, allowing the broadband response of the encoder to be evaluated without pronounced spectral features imposed by a detection crystal. The relatively weak third-order interaction in air also makes this a demanding implementation, as it requires tight focusing and precise probe–THz overlap. Successful detection therefore shows that the encoder preserves sufficient spatial overlap and intensity for this readout. The system is driven by horizontally polarized, approximately 100-fs, 1.2-mJ pulses from a 1-kHz Ti:sapphire amplifier. Approximately 70\% of the pulse energy is used for two-color air-plasma THz generation, while the remaining 30\% forms the optical probe. The THz source follows the configuration reported previously \cite{Ohrt2024BalancedAirBiased}.

\begin{figure*}[!b]
    \centering
    \includegraphics[width=0.95\textwidth]{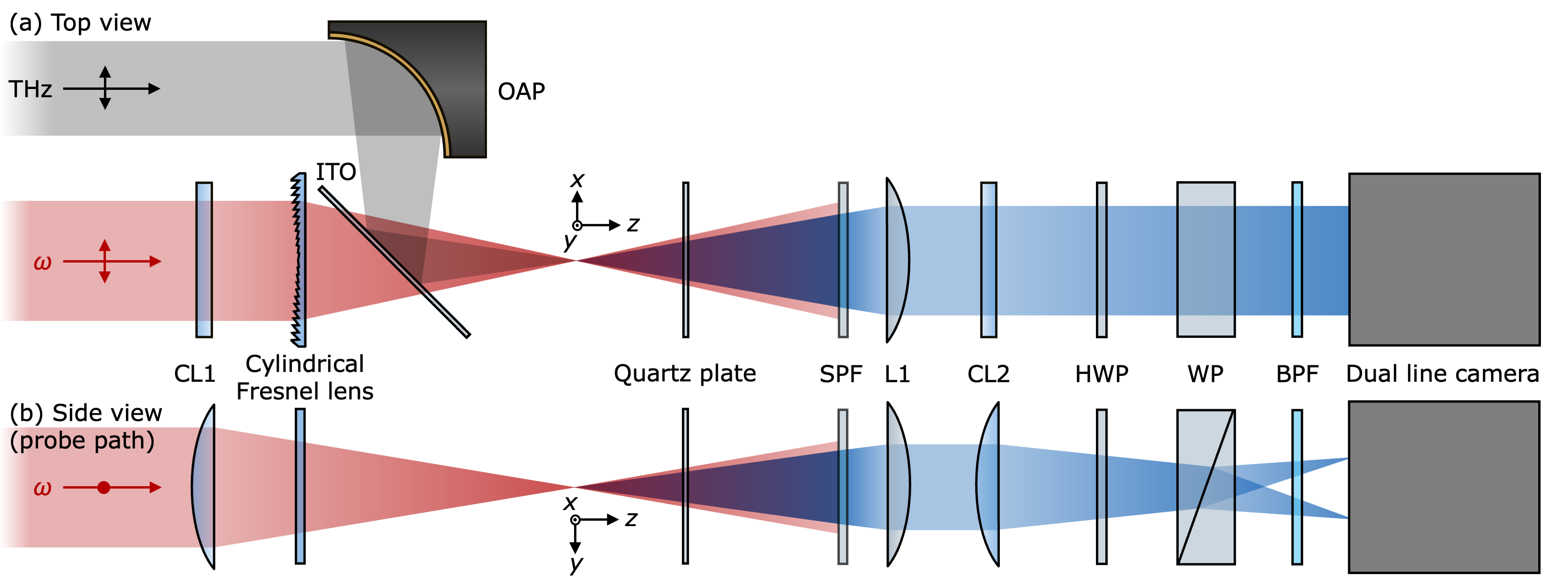}
    \caption{Experimental arrangement for single-shot THz detection using Fresnel lens encoding. (a) Top and (b) side views of the probe and detection paths. The cylindrical Fresnel lens encodes the probe–THz delay along $x$, while cylindrical lens CL1 focuses the probe along $y$. The indium-tin-oxide (ITO)-coated plate combines the probe and THz beams at the detection focus, where terahertz-field-induced second-harmonic (TFISH) radiation is generated. A 45-µm-thick dual-wavelength quartz plate produces the optical local oscillator. The residual probe is removed by the short-pass filter, and the two 400-nm fields are relayed through the balanced polarization readout onto the dual-line camera.}
    \label{fig:3}
\end{figure*}

A 250-mm focal-length lens produces the air-plasma filament, with a 100-µm-thick $\beta$-barium borate crystal positioned approximately 100 mm before the filament to generate the second-harmonic field. A holed 102-mm focal-length off-axis parabolic (OAP) mirror collimates the emitted THz radiation and removes most residual fundamental and second-harmonic light. A high-resistivity silicon wafer at Brewster's angle provides additional optical filtering while preferentially transmitting the horizontally polarized THz field. Two 50.8-mm focal-length OAP mirrors form an intermediate sample focus and recollimate the THz beam. A final 152-mm focal-length OAP mirror focuses the beam into the detection region, as shown in Fig.~\ref{fig:3}(a). An indium-tin-oxide (ITO)-coated glass plate at $45^\circ$ reflects the THz beam and transmits the probe, establishing collinear propagation before the common focus. A schematic of the complete setup, including the THz source and intermediate sample focus, is provided in Supplementary Fig.~S3. Measured far-field and focal-plane THz beam profiles are shown in Supplementary Fig.~S4. An independent conventional EOS measurement at the same focus, using a 30-µm-thick GaP crystal, yielded a conservative peak-field estimate of 300 kV/cm.

The probe passes through a mechanical delay stage and is focused independently along orthogonal transverse axes, as shown in Fig.~\ref{fig:3}. A 150-mm focal-length cylindrical lens (CL1) focuses the probe along $y$ (Fig.~\ref{fig:3}(b)) while the approximately 50-mm focal-length cylindrical Fresnel lens imposes the time-to-space encoding and focuses it along $x$ (Fig.~\ref{fig:3}(a)). Both lenses are positioned approximately one focal length from the interaction region and adjusted to produce a common point focus with the THz beam. The probe--THz interaction at this focus generates THz-field-induced second-harmonic (TFISH) radiation.

After the interaction, the co-propagating probe and TFISH fields pass through a 45-µm-thick crystalline quartz plate, where the probe generates a weak second-harmonic field that serves as an optical local oscillator (LO). The quartz plate acts as a half-wave plate at 800 nm and a full-wave plate at 400 nm. Its azimuthal orientation controls the LO polarization and conversion efficiency while leaving the TFISH polarization unchanged, and its longitudinal position provides additional control of the LO amplitude through the local probe intensity. Before balanced detection, the LO is oriented predominantly orthogonal to the horizontally polarized TFISH field.

A short-pass filter removes the residual 800-nm probe while transmitting the LO and TFISH fields near 400 nm. A 92-mm focal-length lens (L1) collimates these fields, and a 100-mm focal-length cylindrical lens (CL2) focuses the unencoded $y$ coordinate. A half-wave plate at $22.5^\circ$ followed by a Wollaston prism projects both fields into each of two orthogonal polarization channels. A 400-nm bandpass filter with a 10-nm passband provides spectral filtering before the two line-focused channels, denoted A and B, are recorded by a dual-line camera (LineScan-I-Gen2, Glaz; S11639-01 sensor). The camera is triggered synchronously with the 1-kHz laser pulse train and records one dual-line image per shot. A mechanical chopper modulates the THz-generation arm at 500 Hz, producing alternating THz-on and THz-off shots.

The LO--TFISH interference term enters channels A and B with opposite signs. For a given LO field, this term is linear in the TFISH field and therefore proportional to the signed THz electric field. Subtracting the two channels cancels their approximately equal noninterference contributions from the LO and TFISH intensities while adding the field-sensitive interference terms. The corresponding field and intensity relations are derived in section 5 of the Supplementary Material. Any residual LO component projected onto the TFISH polarization produces a DC offset that is removed during post-processing. The arrangement follows the common-mode-rejection principle of balanced air-biased detection \cite{Ohrt2024BalancedAirBiased}, with the electrical bias replaced by a quartz-generated optical LO. It therefore constitutes a quartz-based implementation of balanced optically biased coherent detection \cite{Li2015ControlledOpticalBias,Sun2022BalancedOpticalBias,Huang2025AirAidedCoherentDetection}. Optical bias is used because air breakdown limits the available static field well below the measured THz peak field of approximately 300 kV/cm. Matching a static-bias-generated LO to the TFISH field would therefore require attenuating the THz pulse, reducing the detected 400-nm signal and acquisition signal-to-noise ratio. The quartz plate instead provides a tunable LO that is adjusted to approximately match the TFISH field, maximizing the coherent interference relative to the common intensity background.

\begin{figure*}[!b]
    \centering
    \includegraphics[width=0.95\textwidth]{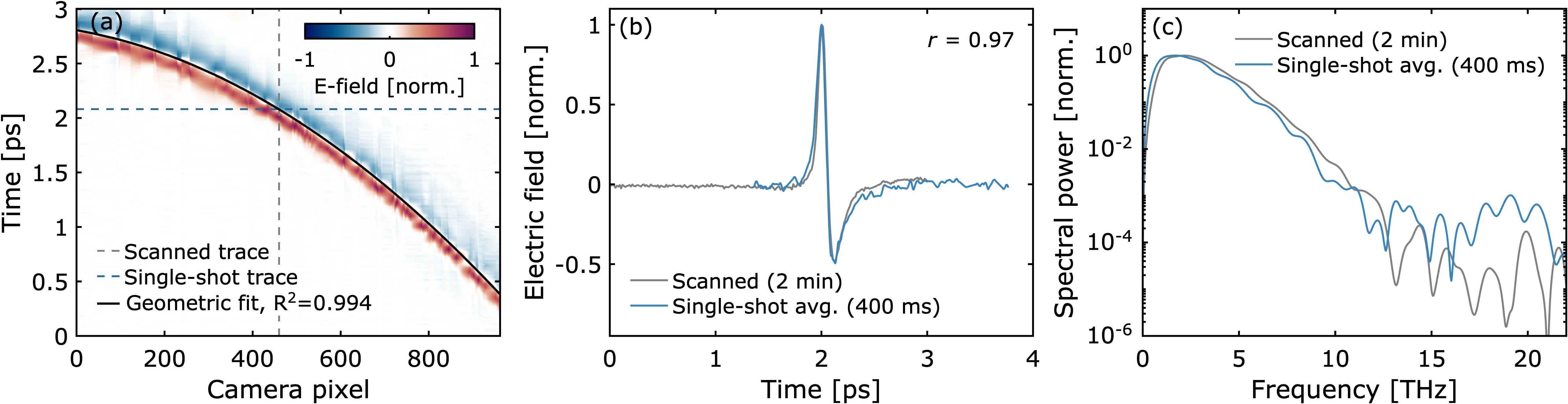}
    \caption{Calibration and validation of the time-to-space mapping. (a) Processed A–B calibration map after retaining pixels with peak normalized cross-correlation above 0.90 and absolute timing residual below 16 fs. The black curve is the fitted geometric mapping ($R^2=0.994$); dashed lines mark the cuts compared in (b). (b) Delay-scanned waveform (gray; 2 min acquisition time) and mean of 200 single-shot THz-on waveforms (blue; 400 ms total, including 200 interleaved background shots). The waveforms are aligned at their principal positive peak, and the correlation coefficient is evaluated over their common temporal interval. (c) Corresponding normalized power spectra calculated over the same interval using identical processing.}
    \label{fig:4}
\end{figure*}

After construction or substantial realignment, a calibration scan determines the relative response of each camera pixel and its corresponding probe--THz delay. At each delay-stage position, alternating THz-on and THz-off images are recorded in both channels and averaged. Background subtraction is performed separately for A and B before calculating the balanced A--B signal. The resulting maps are shown in Supplementary Fig.~S5. As the stage is scanned, the THz signal moves across the camera according to the Fresnel lens time-to-space mapping. Its amplitude varies because of the Gaussian probe profile, with additional spatial modulation from the Fresnel grooves and their diffraction.

Each camera pixel records a conventional delay-scanned waveform. After offset removal and amplitude normalization, each pixel waveform is cross-correlated with a representative scanned waveform, shown in gray in Fig.~\ref{fig:4}(b), and the relative delay at maximum normalized cross-correlation is assigned to that pixel. The resulting arrival times are fitted with the geometric time-to-space mapping in Eq.~(\ref{eq:time_window}). The fit follows the extracted arrival times closely ($R^2=0.994$), supporting the use of the air-path approximation over the experimentally employed interval. Pixels are retained only when their peak correlation exceeds 0.90 and their absolute timing residual from the fit is below 16 fs. These fixed, empirically chosen criteria exclude distorted responses that would otherwise introduce spurious oscillations into the reconstructed waveforms. Such responses are consistent with spatial mixing caused by Fresnel-groove diffraction and imperfect imaging. The fixed mask retains 229 of 1151 camera pixels and is applied unchanged to all subsequent measurements. Each Fresnel groove remains sampled by multiple retained camera pixels. The fitted mapping, shown in Fig.~\ref{fig:4}(a), defines the calibrated, nonuniform time axis used for waveform reconstruction. Supplementary Fig.~S6 presents the processed calibration scan and further spatial and spectral diagnostics; additional processing details are provided in Section 6 of the Supplementary Material. The calibration remains valid as long as the optical alignment is unchanged.

The vertical and horizontal cuts through the same calibration map in Fig.~\ref{fig:4}(a) show a delay-scanned waveform recorded at one pixel and an averaged single-shot waveform recorded at one stage position, respectively. Figure~\ref{fig:4}(b) compares the delay-scanned waveform, acquired in 2 min, with an average of 200 THz-on single-shot waveforms acquired in 400 ms, each paired with an interleaved background frame. For quantitative comparison, the waveforms are aligned at their principal positive peaks and interpolated onto a common temporal grid over their overlapping interval. They have a correlation coefficient of $r=0.97$. Figure~\ref{fig:4}(c) shows the normalized power spectra calculated over the same interval using identical processing. The spectra show similar overall envelopes and both contain spectral content beyond 10 THz. The averaged single-shot spectrum is slightly noisier, resulting in a higher noise floor and visible spectral fluctuations at high frequencies. The measured bandwidth is consistent with previous air-based detection of a similar source driven by 100-fs pulses \cite{Ohrt2024BalancedAirBiased}. Driving pulses of approximately 40 fs have supported detection beyond 30 THz \cite{Ohrt2025ScanningFree}. Accessing this range with Fresnel lens encoding would require a shorter optical probe pulse and may also require smaller local temporal increments across the relevant part of the waveform, attainable using a Fresnel lens with a longer focal length or finer groove spacing. Conversely, narrower-band sources can be matched to a longer observation window and larger temporal increments using a shorter-focal-length or larger-aperture lens and selecting a wider or farther off-axis interval. The lens geometry and illuminated interval can therefore be chosen to match the encoder to the source bandwidth and probe pulse duration.

\subsection{Single-shot ultrabroadband THz-TDS}

\begin{figure*}[!b]
    \centering
    \includegraphics[width=0.95\textwidth]{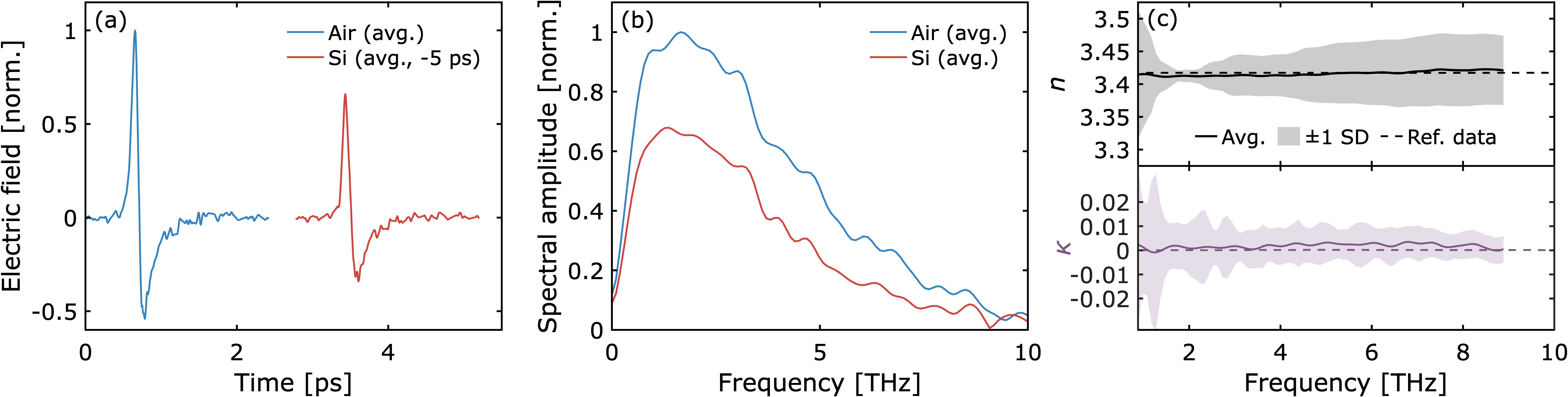}
    \caption{Single-shot ultrabroadband THz-TDS of a 0.964-mm-thick high-resistivity silicon wafer. (a) Averaged electric-field waveforms measured in air and after transmission through silicon. The time origin marks the beginning of the calibrated air-reference window. For compact display, the silicon waveform is shifted by $-5$ ps relative to its measured arrival time. The plotted offset is not used in the material-parameter retrieval. (b) Corresponding spectral amplitudes normalized to the maximum of the air-reference spectrum. (c) Real refractive index $n$ (top) and extinction coefficient $\kappa$ (bottom) retrieved from the averaged waveforms. Shaded regions denote $\pm 1$ standard deviation over 5,000 complete material-parameter retrievals, each using one experimentally measured single-shot reference waveform and one experimentally measured single-shot sample waveform. Dashed curves show 20°C reference data \cite{Franta2017SiliconDispersion}.}
    \label{fig:5}
\end{figure*}

To assess spectroscopic performance, we perform transmission-mode single-shot THz-TDS on a 0.964-mm-thick high-resistivity silicon wafer, Fig.~\ref{fig:5}. Separate ensembles of 200 single-shot waveforms are acquired first in air and then after inserting the wafer. Their averages and corresponding spectral amplitudes are shown in Figs.~\ref{fig:5}(a) and \ref{fig:5}(b), respectively.

The complex refractive index is retrieved from the field transmission $\widetilde{E}_{\mathrm{Si}}/\widetilde{E}_{\mathrm{air}}$ using the standard single-pass relations for a plane-parallel sample \cite{Duvillaret1996MaterialParameters}. The real refractive index $n$ is obtained from the phase of this ratio, while the extinction coefficient $\kappa$ is obtained from its Fresnel-corrected amplitude. The solid curves in Fig.~\ref{fig:5}(c) are calculated from the averaged reference and sample waveforms. Across the 0.8–8.9-THz retrieval range, $n$ remains close to 3.42 and $\kappa$ remains close to zero, consistent with the literature values shown by the dashed curves \cite{Franta2017SiliconDispersion}.

Because the reference and sample ensembles are acquired sequentially, matching waveforms by shot index would be arbitrary. We instead estimate the uncertainty of a single-shot retrieval by independently drawing, with replacement, one waveform from each 200-shot ensemble and repeating the complete retrieval for 5,000 pairings. Each pairing therefore represents a complete material-parameter retrieval from one experimentally measured single-shot reference waveform and one experimentally measured single-shot sample waveform. The shaded regions in Fig.~\ref{fig:5}(c) show $\pm 1$ standard deviation of the resulting single-shot retrievals around the averaged-waveform result. The uncertainty is largest near the low-frequency boundary because both retrieved quantities contain an inverse-frequency dependence and the phase and amplitude become less stable near the edge of the signal band. At higher frequencies, the decreasing spectral amplitude primarily broadens the phase-derived uncertainty in $n$. The $\kappa$ band remains comparatively uniform because its inverse-frequency scaling partly offsets the increasing amplitude noise.

\subsection{Single-shot monitoring of air-plasma perturbations}

To demonstrate shot-to-shot monitoring of a transient perturbation to the THz source, water droplets are released individually from a pipette positioned above the two-color air-plasma filament. Each droplet crosses the filament and abruptly disrupts the THz-generation conditions, providing a visually identifiable event that can be correlated with the simultaneously acquired single-shot waveforms.

\begin{figure}[H]
    \centering
    \includegraphics[width=0.45\textwidth]{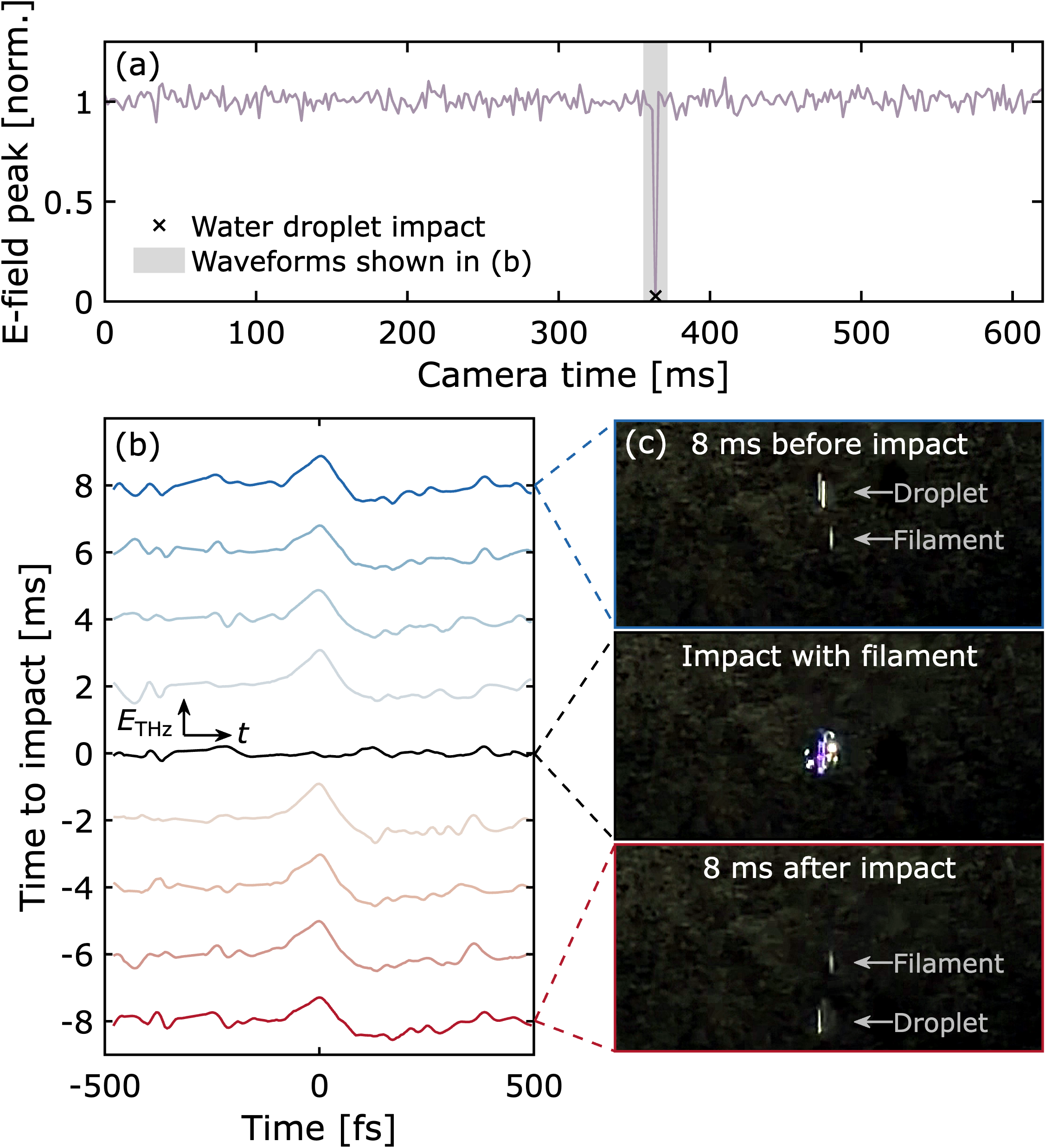}
    \caption{Single-shot monitoring of a water-droplet impact on the air-plasma THz source. (a) Representative 600-ms interval showing the normalized THz field sampled at the temporal position of the principal peak in the median waveform. The cross marks the impact identified from the synchronized video, and the shaded region denotes the interval expanded in (b). (b) THz waveforms acquired at 2-ms intervals from 8 ms before to 8 ms after impact. The traces are vertically offset according to their time relative to impact, with the impact trace shown in black. (c) Synchronized video frames recorded 8 ms before impact, at impact, and 8 ms after impact. Dashed connectors identify the corresponding waveform times in (b).}
    \label{fig:6}
\end{figure}

Calibrated single-shot THz waveforms are acquired continuously at 2-ms intervals, and the field at the temporal position of the principal peak in the median waveform is tracked over time. Figure~\ref{fig:6}(a) shows a representative 600-ms segment, normalized to its median value over the full recording. Excluding the impacts, this field value has a standard deviation of 4.3\% of its mean. At the marked impact, the field falls to nearly zero. Figure~\ref{fig:6}(b) expands the shaded interval and shows that the complete THz waveform is suppressed at impact before recovering in the subsequent acquisitions.

A video camera operating at 120 frames/s records the droplet and filament from above at an oblique angle, synchronized with the THz acquisition. Figure~\ref{fig:6}(c) shows the corresponding frames before, during, and after the selected impact. The droplet is visible primarily through reflected and scattered filament light, while the dark background is the dry air-purged enclosure surrounding the THz beam path. At impact, the bright region indicates that the droplet temporarily surrounds the observed section of the filament. The synchronized images therefore associate the waveform suppression directly with the droplet crossing the plasma.

An extended 5-s recording containing three droplet impacts is shown in Supplementary Fig.~S7. Synchronized video confirms that each field suppression coincides with a droplet crossing the plasma, with all three events showing the same qualitative suppression and recovery.

The experiment is not designed to isolate the physical origin of the suppression or to identify additional THz emission from the liquid. The response is consistent with a combination of disrupted air ionization, scattering of the driving fields, and scattering or absorption of the generated THz radiation by water surrounding the filament. Regardless of the relative contribution of these processes, the measurement demonstrates field-resolved, shot-to-shot monitoring of abrupt perturbations to a fluctuating THz source.

\section{Conclusion}

We have demonstrated Fresnel lens time-to-space encoding for single-shot ultrabroadband THz waveform detection. The calibrated, quadratic mapping spans 2.42 ps with local groove-defined temporal increments of 12--72 fs. Because a single, low-cost component both focuses and encodes the probe, it provides a common interaction focus with sufficient intensity for coherent nonlinear single-shot detection.

Combined with balanced coherent air-based detection, the encoder enables reconstruction of two-color air-plasma THz waveforms with spectral content beyond 10 THz using pulses from a tabletop 1-kHz Ti:sapphire amplifier. Agreement with delay-scanned measurements supports the accuracy of the calibrated nonuniform time axis, while transmission measurements of silicon and synchronized monitoring of droplet-induced source perturbations show spectroscopic and event-resolved operation. Selecting the lens geometry and illuminated aperture allows the balance between local temporal increments and the observation window to be matched to different source bandwidths and probe pulse durations. The encoder could also be combined with electro-optic sampling or coherent detection in thin third-order nonlinear solids. The latter can substantially reduce the required probe energy while supporting ultrabroadband detection \cite{Tomasino2017SolidStateBiased,Zhou2024Chi3Dispersion,Ohrt2025ScanningFree}. Hence, Fresnel lens encoding offers a simple route to field-resolved single-shot THz measurements and may be useful for shot-dependent phenomena for which delay scanning is unsuitable.

\begin{backmatter}

\bmsection{Funding}
Independent Research Fund Denmark (2035-00365B); 
Novo Nordisk Foundation (NNF24OC0094858).
I.H.C. and K.A.N. acknowledge support from the Office of Naval Research under Grant No. 14035670.

\bmsection{Disclosures} The authors declare no conflicts of interest.

\bmsection{Data availability} Data underlying the results presented in this paper are not publicly available at this time but may be obtained from the authors upon reasonable request.

\bmsection{Supplemental document} See the Supplementary Document for supporting content.

\end{backmatter}

\bibliography{references}

\end{document}